\documentclass[12pt]{article}
\usepackage{epsf}
\usepackage{amsmath}
\usepackage{amssymb}
\usepackage{amsfonts}

\newcommand{\uiaddress}
{{\small\it Department of Physics, University of Illinois, Urbana, IL 61801}}
\newcommand{\ucaddress}
{{\small\it Enrico Fermi Institute, University of Chicago, Chicago, IL 60637}}
\newcommand{\email}[1]{\thanks{e-mail: \tt#1}}
\newcommand{\preprint}
{\begin{flushright}\begin{small}
ILL-(TH)-00-07\\ EFI-2000-21\\ hep-th/0006168\\ 
\end{small}\end{flushright}       
}

\newcommand{\eqr}[1]{(\ref{eq:#1})}
\newcommand{\tr}{{\rm tr}}
\newcommand{\qcomm}[2]{\left[#1,#2\right]_q}
\newcommand{\vev}[1]{\langle#1\rangle}

\newcommand{\field}[1]{\mathbb{#1}}
\newcommand{\ZZ}{{\field Z}}
\newcommand{\CC}{{\field C}}
\newcommand{\RR}{{\field R}}

\newcommand{\myfig}[3]{\begin{figure}[ht]
\begin{center}
\leavevmode
\epsfxsize=#2cm
\epsfbox{#1}
\end{center}
\caption{#3}
\label{fig:#1}
\end{figure}}

\begin{document}

\begin{titlepage}
        \title{
        \preprint\vspace{1.5cm}
		Non-Commutative Moduli Spaces,\\ Dielectric Tori and T-duality
		}
		\author{
        	David Berenstein,\email{berenste@pobox.hep.uiuc.edu}
		Vishnu Jejjala,\email{vishnu@pobox.hep.uiuc.edu}\\
		\uiaddress\\
		and\\     
		Robert G. Leigh\email{rgleigh@uiuc.edu}\\
		\ucaddress\\ {\small \it and} \uiaddress
        \\
		}
\maketitle

\begin{abstract}
We review and extend recent work on the application of the non-commutative
geometric framework to an interpretation of the moduli space of vacua of
certain deformations of $N=4$ super Yang-Mills theories. We present a 
simple worldsheet calculation that reproduces the field theory results
and sheds some light on  the dynamics of the D-brane bubbles. Different
regions of moduli space are associated with $D5$-branes of various topologies; 
singularities in the moduli space are associated with topology change. 
$T$-duality on toroidal topologies maps between mirror string realizations of
the field theory.
\end{abstract}
\end{titlepage}

%%%%%%%%%%%%%%%%%%%%%%%%%%%%%%%%%%%%%%%%%%%%%%%%%%%%%%%%%%%%%%%%%%%%%%

\section{Introduction}\label{sec:intro}

The $N=4$ super Yang-Mills theories in four dimensions possess\cite{LS}
exactly marginal deformations
\begin{equation}\label{eq:margsup}
W=\tr \left( \phi_1\phi_2\phi_3-q\phi_2\phi_1\phi_3\right)+
\lambda\tr \left(\phi_1^3+\phi_2^3+\phi_3^3\right)
\end{equation}
The $q$-deformation is of particular interest. In its presence, we find
$F$-term constraints
\begin{equation}\label{eq:constr}
\qcomm{\phi_1}{\phi_2}=0,\ \ \ \
\qcomm{\phi_2}{\phi_3}=0,\ \ \ \ 
\qcomm{\phi_3}{\phi_1}=0
\end{equation}
The solutions of these equations determine the moduli space of 
supersymmetric vacua of the theory. Eqs. \eqr{constr}
give relations in the algebra of $N\times N$ matrices. This algebra 
is non-commutative, and thus the moduli space has a non-commutative 
geometric interpretation. 
Points in this non-commutative geometry are defined to be irreducible 
representations of the algebra \eqr{constr}, up to equivalences (amounting
to gauge transformations on the brane worldvolume). This definition 
is in accord\footnote{In particular, each irreducible representation
${\cal A}\stackrel{\mu}{\to} M_N(\CC)\to0$ provides a maximal ideal, $ker\ \mu$.
Maximal ideals, as in algebraic geometry, correspond to points.}
 with the usual definitions in non-commutative geometry.
This structure persists in general, for the superpotential
\eqr{margsup}, but also for arbitrary relevant single-trace perturbations as well.

In Ref. \cite{BJL1}, we presented a discussion of non-commutative 
moduli space for a variety of deformations. 
Moduli space is built 
of direct sums of irreducible representations of the algebra, and 
given a complete classification of the irreducible representations,
it is possible to describe the moduli space as a non-commutative
version of the symmetric product.

As well, we noted that 
these theories may be obtained from two distinct classes of string
geometries. First, one may obtain these models via $\ZZ_m\times\ZZ_m$
orbifolds with discrete torsion\cite{D,DF,BL}, which is related to 
$q$, for $q^m = 1$. Second, the models are obtained 
through deformations of the 
near-horizon geometry, $AdS_5\times S^5$, dual to the $N=4$ field 
theory. These two models are related by mirror symmetry, which may 
be traced to a T-duality on torus fibrations of the 5-sphere of the 
near-horizon geometry. The significance of this torus manifests 
itself when one considers $D3$-branes in background $NS$ and $RR$ 
fields.  We will describe these phenomena in detail below.

The principal effect here is that $D3$-branes in these backgrounds
become $D5$-branes of non-trivial topology. This phenomenon has 
appeared in several guises over the past few years. First, it
appeared in Ref. \cite{BFSS} wherein membranes of Matrix theory were
understood to be built from $D0$-branes. Spherical membranes in Matrix
theory were considered in Ref. \cite{KTay,BC}.
More
recently, Myers\cite{RM} has shown that $Dp$-branes in $RR$ 
backgrounds in string theory carry a $D(p+2)$-brane dipole 
moment, and thus may be thought of as $D(p+2)$-branes with 
topology $\RR^p\times S^2$. This effect comes into play, for example,
for relevant
deformations\cite{PS}  within the $AdS/CFT$ correspondence.
One of the difficult issues is to 
understand this phenomenon for a single $Dp$-brane, as then the
non-commutative geometry that one obtains is trivial, and one can
just as well regard the brane as pointlike, as it seems to have no structure.
Moreover, in this regime the $\alpha'$
corrections are not suppressed by $1/N$, and the DBI action is also 
unreliable.

One purpose of this letter is to try to understand this phenomenon of
brane bubbling into spheres for a 
single $Dp$-brane by using worldsheet methods. 
We begin with a detailed account of the non-commutative
moduli space of vacua. We then consider the AdS/CFT realization of 
these field theories. Here we are interested in a probe calculation,
and so small deformations in the geometry can be taken into account 
systematically using worldsheet methods. 
This fills gaps in the discussion of Ref. \cite{BJL1}:
the existence of certain massless modes can be proven and is important
for the realization of the non-commutative moduli space.
$Dp$-branes may be thought
of as $D(p+2)$-branes of various topologies. There is a T-dual orbifold
description; singularities in the latter description, such as the fractionation
of branes, corresponds directly to the degenerations of these
topologies.

\section{General Features of the Moduli Space}

Let us begin with a description of the moduli space for the 
$q$-deformed theory, where $q^n=1$ for some integer $n$. When
$q$ is such a root of unity, there are branches of moduli space
which do not otherwise exist.
First, we note that the center of the algebra
is generated by the elements $x=\phi_1^n$, $y=\phi_2^n$, $z=\phi_3^n$
and $w=\phi_1\phi_2\phi_3$, and that these satisfy the matrix
relation
\begin{equation}
(-w)^n+xyz=0
\end{equation}
Thus the center of the algebra reproduces the commutative moduli 
space, the orbifold upon which we expect point-like 
$D$-branes to propagate. We will use orbifold language to describe
the moduli space here, but there is also a mirror description
which we will give details of in the next section.
The orbifold space has singularities along
complex lines where two of the $x,y,z$ vanish.
\myfig{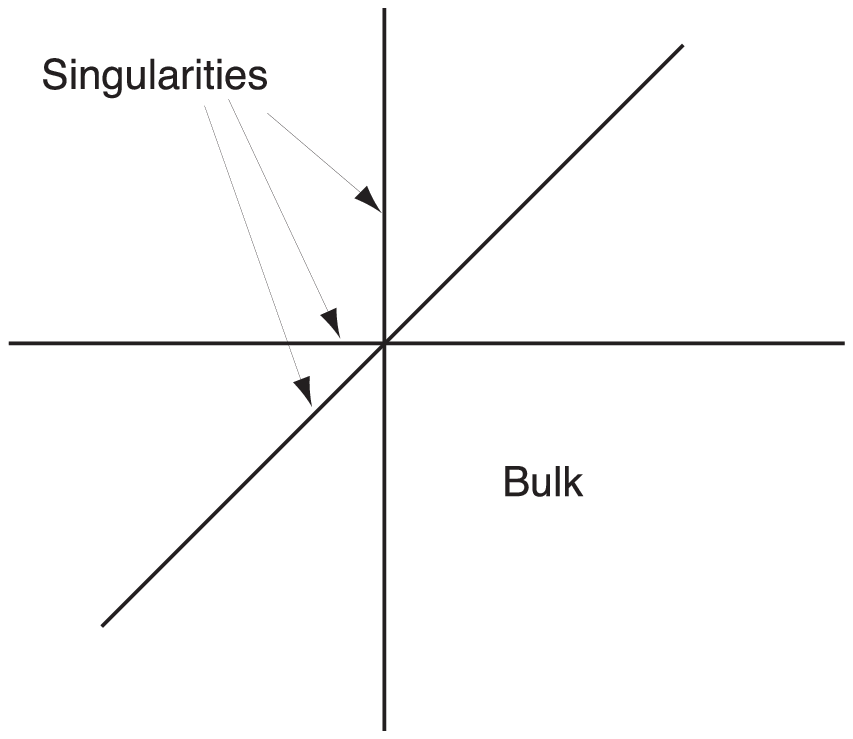}{4}{Moduli space of $q$-deformed theory.}

Next, we look for representations of the algebra
\eqr{constr}. In the bulk of the orbifold, where at least
two of the $x,y,z$ do not vanish, we find an $n$-dimensional\footnote{We
assume that $n\le N$.} representation\footnote{This representation has
been chosen so that $\phi_3$ is diagonal at the singularity $x=y=0$. At
other singularities, one may parameterize in a different way. This may
be thought of in terms of patches.} $R(a,b,c)$
\begin{eqnarray}
\phi_1&=&aQ\\
\phi_2&=&bQ^{-1}P^{-1}\\
\phi_3&=&cP
\end{eqnarray} 
where $a,b,c$ are arbitrary complex numbers and $P,Q$
satisfy $\qcomm{P}{Q}=0$. An explicit representation is
\begin{equation}\label{eq:PQ}
P = \begin{pmatrix}1&0&0 &\dots& 0 \\ 0& q & 0
& \dots&0 \\ 0& 0& q^2&\dots&0 \\
\vdots&\vdots&\vdots& \ddots&\vdots \\
0&0&0&\dots&
 q^{n-1}\end{pmatrix},\ \ \ \
Q = \begin{pmatrix}0&0& \dots&0&1\\
1&0&\dots&0&0 \\
0&1&\dots&0&0 \\
\vdots&\vdots&\vdots&\ddots&\vdots\\
0&0&\dots&1&0
\end{pmatrix}
\end{equation}

The individual (diagonal) elements of $P$ represent fractional branes, as
can be seen by going to the singularities. Indeed, at 
for example $x=y=0$, we find $n$ one dimensional representations
$R(0,0,c)$, $R(0,0,qc)$ $\ldots,$ $R(0,0,q^{n-1}c)$ which are all
distinct. As one approaches the singularity from the bulk, 
the bulk representation
$R(a,b,c)$ becomes reducible, and decomposes into the direct sum
\begin{equation}\label{eq:decompose}
\lim_{a, b\to 0} R(a,b,c) = 
R(0,0,c)\oplus R(0,0,qc)\oplus\dots R(0,0,q^{n-1}c)
\end{equation}
For $n<N$, we can build representations by composing these $n$-dimensional 
representations, together perhaps with one-dimensional 
representations associated to the fractional branes.
When $q$ is not a root of unity, the representations
that we have given do not exist, and the moduli space is reduced
considerably, so only the fractional branes survive.

The decomposition (\ref{eq:decompose}) may be represented graphically as
in Fig. \ref{fig:quiver}. Note that this diagram is precisely the quiver
diagram for the local singularity. Each node in the quiver represents one
of the one-dimensional representations, and the lines connecting them represent
the vevs of fields which vanish at the singularity.
\myfig{quiver.epsf}{4}{Quiver diagram for reducible representations.}\label{fig:quiver}

\subsection{Relevant Deformations}

It is interesting to see what happens to this moduli space when
relevant deformations are added. We consider first the case of
a rank one mass term $W=\frac{1}{2}m\phi_3^2$. When $q=1$, it is
clear what happens: at scales below $m$, we integrate out the 
field $\phi_3$, and the dimension of the moduli space is reduced
from $3N$ to $2N$. This is also true for $q=-1$. 
For $q^n=1, n>2$, the mass term does not lift
the moduli space in this way. Instead, it deforms it to\footnote{The
Casimir $w$ is shifted to 
$w=\phi_1\phi_2\phi_3+t\phi_3^2$.}
\begin{equation}\label{eq:onemassv}
(-w)^n+xyz=-t^n z^2
\end{equation}
where $t=m/(1-q^2)$.
The representation $R(a,b,c)$ is also suitably deformed 
to
\begin{eqnarray}
\phi_1&=&aQ\\
\phi_2&=&bQ^{-1}P^{-1}+dQ^{-1}P\\
\phi_3&=&cP
\end{eqnarray} 
where $ad(q^2-1)=mc$. Thus, although $\phi_3$ is determined, 
there is a third independent complex parameter -- the moduli
space has not decreased in dimension by the addition of the
mass term. 
Note that one can solve for $\phi_3$ by the $F$-term
equations and thus ``integrate'' it out of the superpotential. 
However, this procedure is misleading in this case, because
$\phi_1,\phi_2$ have more degrees of freedom than one na\"\i vely
expects.

The mass term does resolve one of the singular complex lines;
indeed, the variety \eqr{onemassv} is singular at $x=w=z=0$ and 
$y=w=z=0$. Along the line $x=y=0$, the representation is no
longer reducible, as it was for $m=0$. 
\myfig{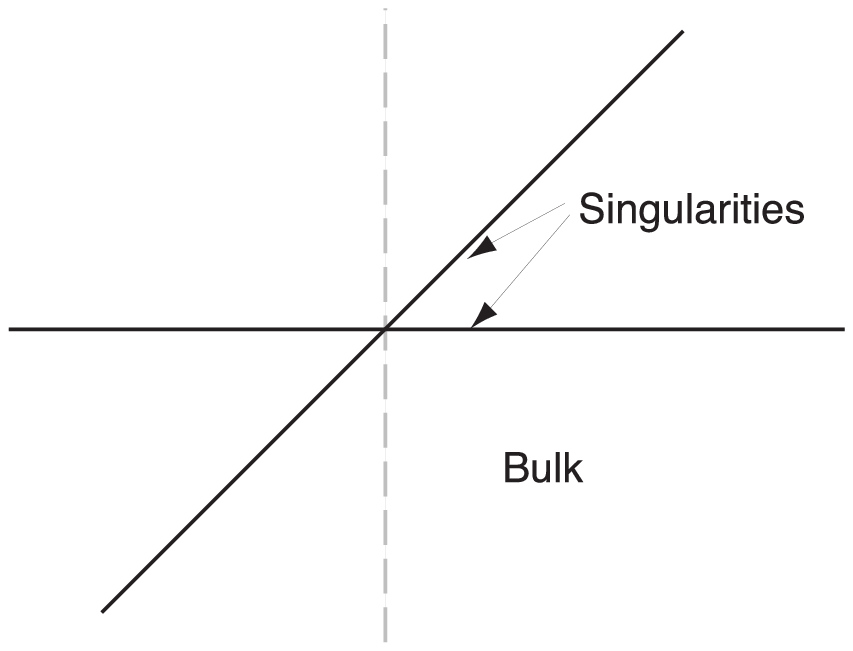}{4}{Moduli space of $q$-deformed theory, with rank one
mass.}

One may also consider higher rank mass terms. One such rank three
mass term may be written (when $q\neq1$), through a 
holomorphic field 
redefinition, as
\begin{equation}
W=\tr\left( \phi_1\phi_2\phi_3 - q\phi_2\phi_1\phi_3 +
\frac m2\phi_3^2+\zeta_3\phi_3\right)
\end{equation}
This further deforms the Casimir $w$, resulting in a deformed 
commutative geometry
\begin{equation}
xyz +(-w)^n = - t^n z^2+ t_\zeta^n z
\end{equation}
where $t_\zeta=\frac{\zeta_3}{q-1}$. Again, the 
irreducible representation 
has three independent complex parameters
\begin{eqnarray}
\phi_1&=&aQ\\
\phi_2&=&bQ^{-1}P^{-1}+dQ^{-1}P+eQ^{-1}\\
\phi_3&=&cP
\end{eqnarray} 
where $ad(q^2-1)=mc$ and $ae(q-1)=\zeta_3$. 
The two complex lines of singularities that were present for $\zeta_3=0$
are smoothed out to the hyperboloid
\begin{equation}\label{eq:sing3m}
xy=t_\zeta^n,\ \ \ z=w=0
\end{equation}
Along this singularity, one can take the $n$th root of eq.\eqr{sing3m},
and obtain one-dimensional representations (corresponding to fractional
branes).
\myfig{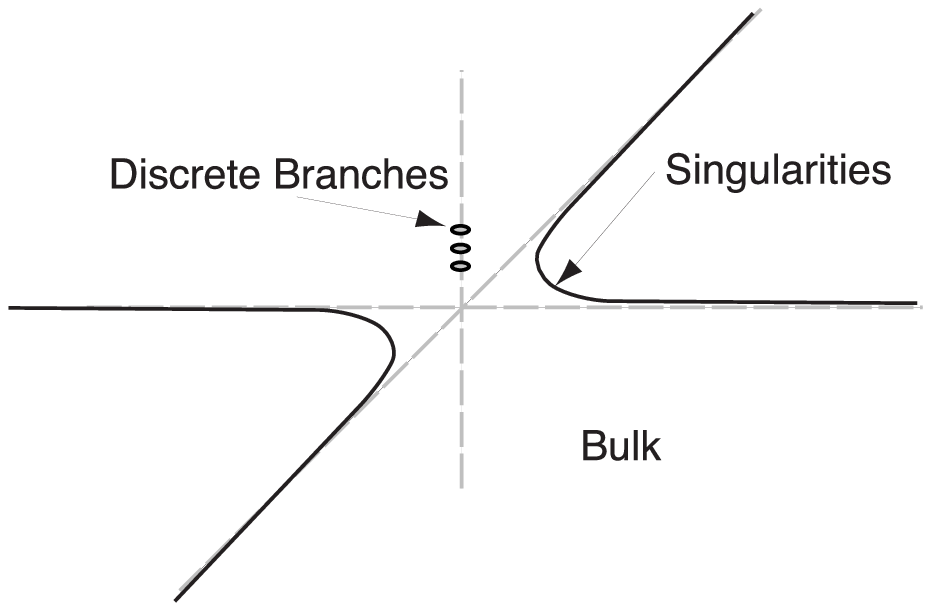}{5}{Moduli space of $q$-deformed theory, with rank three
mass.}

In addition, there are isolated representations that are not
covered by the above solution. For example, there are representations
for which $\phi_1^n=0$ non-trivially. This can happen for nilpotent $\phi_1$
where $\phi_1^k=0$ for any $1\le k\le n$. For $k<n$, these representations may be
thought of as $q$-deformed $SL(2)$ representations, as we may bring
the algebra to the form (for $q\neq 0,1$).
\begin{eqnarray}
\qcomm{A_+}{A_-}&=&2A_0, \label{eq:algo}\\
\qcomm{A_0}{A_+}&=&A_+,\\
\qcomm{A_-}{A_0}&=&A_-\label{eq:algt}
\end{eqnarray}
Each of these $n-1$ special branches occur at a fixed
value of $z$ along $x=y=0$. 

The rank three mass deformation that we have considered here is special,
in that it is equivalent to mass eigenvalues $(m,-m,m')$. Other rank
three mass terms lead to algebras which are different from 
\eqr{algo}--\eqr{algt} as one cannot diagonalize the mass terms without
altering the form of the $q$-commutator. Nevertheless, the physics is
very similar\cite{BJL1} to the case discussed here.

\section{Spheres and Tori}

Next, we consider the moduli space of these field theories by
looking at the dual near-horizon string theory. The orbifold 
field theory has a description in terms of the near-horizon
geometry $AdS_5\times S_5/\Gamma$ with discrete torsion encoded
in the boundary conditions of massless twisted states. Here we
are interested instead with the deformation of the $N=4$ theory.
In the AdS/CFT correspondence\cite{M,W,GKP} we are instructed to
interpret deformations of a superpotential in terms of 
non-normalizable modes of supergravity states. The $q$-deformation
is induced\cite{KRN} by elements of the ${\bf 45}$ of $SU(4)_R$ which appears
as the second harmonic of the anti-symmetric tensor modes,
$G_{(3)}=F_{(3)}-\tau H_{(3)}$. 
As the deformation
preserves conformal invariance (indeed, $G_{(3)}$ is independent of
the $AdS_5$ radial coordinate), the geometry is of the form 
$AdS_5\times {\cal N}$ for some ${\cal N}$. 

As we have reviewed earlier, $D$-branes in background $RR$ fields 
attain multipole moments of branes of higher dimension. 
Fractional $D3$-branes should be thought of as
$D5$-branes of topology $\RR^4\times S^2$, oriented in a certain
way.
There are actually two effects in the present case. 
First, the 5-brane is electrically charged
with respect to $F_{(7)}\sim *_4 \tilde F_{(3)}$, 
which has support on $\RR^4\times D^3$,
where $D^3$ is bounded by the $S^2$. This field is responsible 
for the di-electric effect. Secondly, there is a
background $H_{(3)}^{NS}$. This is an $AdS_5$ scalar, and so has all
three indices along ${\cal N}$. Thus, $H_{(3)}$ contributes to the energy
of a 5-brane if the $D^3$ has components along ${\cal N}$. Taking both 
fields into account, we then expect that the radius of the disc $D^3$
is oriented partially in the $AdS_5$ direction, and partially along 
${\cal N}$. As shown in Ref. \cite{BJL1}, energy considerations
from the DBI action suggest, when the deformation is small, 
that the 2-sphere is of size
\begin{equation}
\vev{r}\sim (c_{NS}+g_{str}c_R)(4\pi^2\alpha')^2
\end{equation}
where $\int_{D^3}H_{(3)}=c_{NS}r^3$, $\int_{D^3}\tilde F_{(3)}=c_{R}r^3$.
Thus the 2-spheres are expected to be of string size, and one should be hesitant
in accepting the geometric picture described here. Nevertheless, we will continue
with this way of speaking, as it provides useful intuition for the applications
studied here. 

We are interested in the new branches of moduli space which appear 
near the singularities. Near such a singularity, when $q^n=1$, the 
background which corresponds to the 
$q$-deformation consists of $H_{(3)}$ only. 
As a result, the di-electric
effect vanishes due to the orientation of the background field.

Small $D5$-branes with spherical topology correspond to fractional branes in
the orbifold dual. If the spheres intersect each other, then 
semi-classically we expect that there are massless string modes stretching 
between
them. However, the radii of the 2-spheres are small in string units, and
thus this intuition may not necessarily be reliable. Moreover, two spheres 
coming
together have locally opposite orientations, and thus correspond to a 
brane-antibrane system where one expects complicated dynamics.
In the following subsection,
we will show using a worldsheet computation that massless modes are 
present if we
tune the separation of $D3$-branes appropriately. This occurs at finite
separation in the presence of a non-zero $G_{(3)}$ field.
Finally, in Section 4, we will argue that the 
existence of these massless modes gives
full consistency with field theory expectations. 

\subsection{Massless States}

We will demonstrate the existence of massless states between 5-brane spheres by
returning to the description in terms of $D3$-branes in weak backgrounds.
In particular, we will show that massless states arise when the $D3$-branes are
held at fixed non-zero separation. This will be done at lowest order in the 
background expansion, but it is clear that the massless states persist in 
stronger backgrounds. 

Note that, from the field theory perspective, we expect to see
new massless modes when the branes are at special
locations. Indeed, from the $q$-deformed superpotential, 
there are 
off-diagonal masses (for $\phi_3={\rm diag}(a_1,a_2)$) 
proportional to $a_1-qa_2$ and $a_2-qa_1$. Thus, massless states are
present for $|a_1|=|a_2|$ if we have
$|q|=1$. It is only when $q$ is a root of unity that the new branch in moduli
space can open up, but the massless states between two $D$-branes persist.

In Ref. \cite{BJL1}, we
simply assumed that these massless states are present because of field theory 
considerations. 
There are several ways in which this effect may
be seen from worldsheet computations. Here, we will
consider the equation of motion 
of an open-string fermion. As is well-known, the Dirac equation is obtained
by integrating the BRST current around the vertex operator. In this case, 
the background makes its presence felt through contact terms with the 
BRST operator.\myfig{brstloop.epsf}{3}{Disc diagram for fermion EOM.}
At lowest order in the $RR$ background, 
the contact terms have been worked out in Ref. \cite{LBQ,LB2}; 
the $NSNS$ contact
terms are standard. We will denote the separation of the branes by a
vector $a^\mu$, of length $a$ and direction orthogonal to the branes. There is
of course a linear contribution in $a^\mu$ to the mass of the stretched string
state. The Dirac equation then takes the form
\begin{equation}\label{eq:Dirac}
\left(i\nabla\cdot\Gamma+\frac{1}{2\pi\alpha'}a\cdot\Gamma+
H_{\mu\nu\rho}\Gamma^{\mu\nu\rho}
+gF_{\mu\nu\rho}\Gamma_{||}\Gamma^{\mu\nu\rho}\right)\psi=0
\end{equation}
The fields $H_{(3)}$ and $F_{(3)}$ have identical orientations in the 
background corresponding to the $q$-deformation. In an appropriate 
coordinate system near the singularity
$x=y=0$, $G_{(3)}$ is oriented along $d\theta\wedge 
(d\bar x \wedge dx-d\bar y\wedge dy)$ and $H_{(3)}/g$ ($F_{(3)}$)
is the imaginary (real) part. It is then easy to see that 
the $\Gamma$-matrix structure in eq. \eqr{Dirac} has half of its
eigenvalues equal to zero when $a=0$. Also, by aligning $a^\mu$
appropriately, here along $z$, there are zero eigenvalues for a 
particular value of $|a|$ proportional to the field strength. This 
corresponds to a chiral fermion in four dimensions. The value of
$|a|$ at which this occurs agrees with the Born-Infeld calculation
presented above (up to vertex normalizations that we have not considered
carefully). 

\subsection{Bulk branes and Tori}

Given the presence of massless modes at the intersections of 2-spheres,
it is natural to ask under what conditions one can turn on vevs for these
modes. These vevs would have an interpretation in terms of smoothing the
singularity of joined spheres. However, by consideration of appropriate
string diagrams (or the equivalent field theory superpotential), we see
that there is a potential preventing this condensation in general, consistent
with supersymmetry. There is however a special configuration, corresponding
to a linear combination of blowup modes, which evades this potential. This
occurs when we have $n$ 2-spheres joined and wrapping around the 5-sphere,
as is suggested by Fig. \ref{fig:spheres.epsf}.
\myfig{spheres.epsf}{5}{The $n$-pinched torus as a string of fractional 
5-brane spheres.} This should be compared to the quiver diagram of
Fig. \ref{fig:quiver}. Turning on the vev referred to above corresponds to 
smoothing the structure in Fig. \ref{fig:spheres.epsf} out to a 2-torus, and is equivalent, in the
orbifold language, to the process
of moving a brane off of the singularity into the bulk. Thus a bulk
3-brane should be thought of as a $D5$-brane with topology $\RR^4\times T^2$.
The space ${\cal N}$ (the deformed $S^5$) should then be thought of 
topologically in terms of a torus fibration over $S^3$.
Large tori are governed
by the Born-Infeld action and are extrema of this action.
Indeed, one can show that near the degeneration, the Born-Infeld action is
independent of the size of the torus; to see this, one needs only the first-order
analysis.

This description is related to the orbifold theory 
by T-duality. 
The T-duality extends in a natural way to non-commutative geometry. 
On one hand,
we are instructed to think of an irreducible representation of the 
non-commutative
algebra as a point in a non-commutative (moduli) space\cite{BJL1}. On the other
hand, we often think of this very same algebra as that of the non-commutative
torus, as in \cite{BFSS}. 
There is no tension between these two concepts;
they are just mirror to each other. 

\subsection{Degenerations}

To further explore the duality between the two descriptions, 
let us consider from a geometric standpoint
the various degenerations of the toroidal 5-branes, and
how these processes are related to singularities in the orbifold 
moduli space. In the massless case, as we have stated, the degenerations
in the moduli space correspond directly to brane fractionation.

With a rank one mass term, one of the singular lines is removed; algebraically,
the representation of the algebra is no longer reducible there. Topologically,
the torus pinches at those points in the moduli space, but only once, so
no moduli are available
to separate the surface. 
On a torus fibration, this fiber is singular and becomes
a sphere with two points identified.

With a rank three mass term, the singularity is a hyperboloid,
and again, along this locus, the torus degenerates to a collection of spheres.
When a brane approaches the locus of the special isolated
branches (at fixed $z$, $x=y=0$), it may be able to split 
into different localized branes, although we have not
been able to check this assertion precisely. In this case, we would get a
degeneration into a collection of spheres, but no new moduli.

One can in fact interpret the matrix representations topologically. Diagonal
entries, such as in the matrix $P$, correspond to spheres, while non-zero
off-diagonal elements can be thought of in terms of a tube joining different
spheres. The genus of the resulting Riemann surface is encoded in the 
number of off-diagonal elements of the generators. Degenerations may be 
understood in this language in terms of the vanishing of matrix elements.

\section{Final remarks and outlook}

In this letter we have seen various features of the interactions between
$D$-branes and weakly deformed backgrounds. The semi-classical 
configuration of a $D(p+2)$-brane wrapping an $S^2$ provides us with
useful intuition and is consistent with worldsheet results. In a
sense however, the branes are truly pointlike, in that the spheres may
pass through each other without deformation. 

The effect of the background on the world-sheet is quantum-mechanical, and
presumably may be thought of in terms of a shift in 
the zero point energy and modings of stretched strings.

From a dual perspective,
the masslessness of these states arises because the fractional branes
are coincident.
The relation between these two different points of view, when the geometry
can be well understood, corresponds to $T$-duality. 
Such dualities are expected to persist to more general situations.

\bigskip
\noindent {\bf Acknowledgments:} We wish to thank M. Strassler and 
A. Hashimoto for discussions. RGL thanks the Enrico Fermi Institute at the
University of Chicago for hospitality while much of this work was carried out.
Work supported in part by U.S.
Department of Energy, grant DE-FG02-91ER40677 and 
an Outstanding Junior Investigator Award. 

%\bibliography{deformation} \bibliographystyle{uiuchept}
\providecommand{\href}[2]{#2}\begingroup\raggedright\endgroup

\end{document}